\documentclass[12pt]{article}
\usepackage{graphics,cite,amssymb,epsfig,float,psfrag,axodraw}
\usepackage[usenames,dvips]{color}
\usepackage{rotating}

\begin{document}

\baselineskip=17pt

\newcommand{\lsim}{\raisebox{-0.13cm}{~\shortstack{$<$ \\[-0.07cm] $\sim$}}~}
\newcommand{\gsim}{\raisebox{-0.13cm}{~\shortstack{$>$ \\[-0.07cm] $\sim$}}~}
\newcommand{\ra}{\rightarrow}
\newcommand{\ee}{e^+e^-}
\newcommand{\tb}{\tan \beta}
\newcommand{\s}{\smallskip}
\newcommand{\nn}{\noindent}
\newcommand{\non}{\nonumber}
\newcommand{\beq}{\begin{eqnarray}}
\newcommand{\eeq}{\end{eqnarray}}

\begin{flushright}
MPP--2003--35\\
CERN TH/2003--161\\
PM/03--16\\
July 2003
\end{flushright}

\vspace*{1cm}

\begin{center}

{\large\sc {\bf NNLO QCD corrections to the Higgs-strahlung}}

\vspace*{3mm}

{\large\sc {\bf processes at hadron colliders}}

\vspace*{7mm}

{\sc Oliver BREIN$^1$, Abdelhak DJOUADI}$^{2,3}$ and {\sc Robert 
HARLANDER}$^{2}$ 
\vspace*{7mm} 

$^1$ Max-Planck-Institut f\"ur Physik, F\"ohringer Ring 6, D--80805 Munich,
Germany. 
\vspace*{2mm}

$^2$ Theory  Division, CERN, CH--1211 Geneva 23, Switzerland.
\vspace*{2mm}

$^3$ Laboratoire de Physique Math\'ematique et Th\'eorique, UMR5825--CNRS,\\
Universit\'e de Montpellier II, F--34095 Montpellier Cedex 5, France. 
\end{center} 

\vspace*{1cm} 

\begin{abstract}

\nn We implement, at next-to-next-to-leading order, the QCD corrections to
Standard Model Higgs  boson production in association with vector bosons at
hadron colliders, $q\bar{q} \to HV$ with $V=W,Z$. They consist of the two--loop
corrections to the Drell--Yan  process for the production of off-shell vector
bosons, $q\bar{q} \to V^*$, and in the case of $Z$ final states, of the
additional contribution from heavy-quark loop mediated processes, in particular
$gg \to HZ$. For the Higgs boson masses relevant at the  Tevatron and the LHC,
$M_H \lsim 200$--300 GeV, the two-loop corrections are small, increasing the
production cross sections by less than 5\% and 10\%, respectively; the scale
dependence is  reduced to a level of less than a few per cent. This places
these processes among the most theoretically clean Higgs boson production
channels at hadron colliders. 
\end{abstract}

\newpage

\subsection*{1. Introduction} 

The Standard Model (SM) predicts the existence of a scalar particle, the Higgs
boson,  that is the remnant of the electroweak symmetry-breaking mechanism that
generates  the weak gauge boson and fermion masses \cite{HHG}. The search for
this particle is the primary mission of  present and future  high-energy
colliders. The Higgs boson can be discovered at Run II of the Tevatron if  it
is relatively light, $M_H \lsim 200$ GeV, as suggested by the fits to the
high-precision electroweak data \cite{EWdata}, and if sufficient integrated 
luminosity is collected \cite{TeV,Houches}. Higgs bosons with masses up to 
$M_H \sim 1$ TeV, a value beyond which perturbation theory is jeopardized in
the SM, can be probed at the upcoming LHC \cite{LHC,Houches}. \s

One of the most important Higgs boson production mechanisms at hadron colliders
is the Higgs-strahlung process, i.e. the associated production of Higgs and
weak gauge bosons, $q \bar{q} \to HV$ with $V=W,Z$ \cite{HVLO}. At the
Tevatron, Higgs particles can be mainly produced in the channel $q\bar{q} \to
HW$ with the $W$ boson decaying into $\ell \nu$ pairs [with $\ell = e, \mu]$
and the  Higgs boson decaying into $b\bar{b}$ or $W^+ W^-$ pairs \cite{HVTev}. 
At the LHC, a plethora of production channels can be used to search for the
Higgs particle; one of the principal detection modes is expected to be the
gluon--gluon fusion process, $gg \to H$ \cite{ggH,ggHNNLO}, with the signatures
$H \to \gamma \gamma$ or $H \to ZZ^{(*)}, WW \to 4\ell$ in, respectively, the
low and high Higgs mass ranges. However, although with more difficulty, the
Higgs boson can also be detected through the channels $q\bar{q} \to HW/HZ$, in
particular in the $\gamma \gamma$ plus lepton final states
\cite{HVLHC,DYequiv}.  These processes could play a  very important role in
the  determination of the Higgs boson properties \cite{Zep}. \s

It is well known that, in hadronic collisions, the lowest-order (LO) cross
sections are affected by large uncertainties arising from higher-order QCD
corrections. If at least the next-to-leading order (NLO) radiative
corrections are  included, the cross sections can be defined properly and their
unphysical variation with the scales are stabilized. To have an even better
control on  the theoretical prediction, the next-to-next-to-leading order
(NNLO) corrections, which are in general very complicated to calculate, are 
desirable. Up to now, the NNLO corrections to SM Higgs boson production at
hadron colliders are known  only for the $gg \to H$ mechanism \cite{ggHNNLO} in
the infinite top quark mass limit\footnote{The NNLO corrections to the
bottom-quark fusion mechanism, $b\bar{b} \to H+X$, which plays an important
role in supersymmetric extensions of the SM, have also been derived recently
\cite{bbH}.}. \s

In this paper, we will discuss the NNLO order, i.e. the ${\cal O}(\alpha_s^2)$,
corrections to the $pp \to HV$ production cross sections [hereafter, we will
use the notation $pp$ for both $pp$ and $p\bar{p}$]. Part of these corrections
are simply those of the Drell--Yan process $pp \to V$;  however, in the case of
$pp \to HZ$, additional corrections are due to diagrams involving the $Zgg$
vertex as well as to  loop-induced $gg$ fusion diagrams.  We show that while
the NNLO corrections increase the  cross sections by only 5 to 10\%, the scale
dependence of the latter is drastically reduced, making these channels
among the theoretically cleanest Higgs production mechanisms. \s

The paper is organized as follows. In the next section, we  describe the known
behaviour of the production cross section at leading and next-to-leading orders
in QCD. In section 3, we summarize the main contributions to the cross section
at NNLO: the Drell--Yan corrections and the additional corrections to $HZ$
production due to gluon--gluon-initiated processes.  The numerical results for
the $K$-factors, the variation with the renormalization and factorization
scales are summarized in section 4.  A short conclusion and some remarks on the
remaining uncertainties on the production cross section are given in the last
section. 

\subsection*{2. LO and NLO cross sections}

The associated production of Higgs and gauge bosons, Fig.~1, is one of the
simplest production mechanisms at hadron colliders: the final state does not
feel strong interactions, which affect only the quark and antiquark initial 
state.  In fact, this process can be viewed simply as the Drell--Yan 
production of a  virtual $W$ or $Z$ boson, which then splits into a real
vector boson and a Higgs particle. Denoting by $k$ the momentum of the virtual
gauge boson, the energy distribution of the full subprocess can be written 
at leading order as
\beq
\frac{ {\rm d} \hat{\sigma} }{ {\rm d}k^2 } (q\bar{q} \to H V) =  \sigma
(q\bar{q} \to V^*) \times \frac{ {\rm d} \Gamma }{ {\rm d}k^2 }(V^* \to H V)\ ,
\eeq
where, in terms of $0\leq k^2\leq Q^2=\hat{s}$ with $\hat{s}$ the
centre-of-mass  energy of the subprocess and the usual two-body phase-space
function $\lambda(x,y;z)$ =$(1-x/z- y/z) ^2-4xy/z^2$, one has
\beq 
\frac{ {\rm d} \Gamma }{ {\rm d}k^2 } (V^* \to H V) = \frac{ G_F M_V^4}{
2\sqrt{2} \pi^2}  \frac{\lambda^{1/2} (M_V^2, M_H^2; k^2)}{(k^2-M_V^2)^2}
\left(1 + \frac{\lambda(M_V^2, M_H^2;k^2)}{12M_V^2/k^2} \right) \ .
\eeq
\begin{center}
\SetWidth{1.1}
\hspace*{4cm}
\begin{picture}(300,100)(0,0)
\ArrowLine(0,25)(40,50)
\ArrowLine(0,75)(40,50)
\Photon(40,50)(90,50){4}{5.5}
\DashLine(90,50)(130,25){4}
\Photon(90,50)(130,75){4}{5.5}
\Text(-10,20)[]{$q$}
\Text(-10,80)[]{$\bar{q}$}
\Text(60,65)[]{$V^* (k)$}
\Text(139,20)[]{$H$}
\Text(159,80)[]{$V=W,Z$}
\end{picture}
\vspace*{-8mm}
\end{center}
\centerline{Figure 1: \it Diagram for associated Higgs and vector boson
production in hadronic collisions.} 
\vspace*{3mm}

The total cross section for the subprocess is obtained by integrating over 
$k^2$\,: 
\beq
\hat{\sigma}_{\rm LO}(q\bar{q} \ra V H)= \frac{G_F^2 M_V^4}{288 \pi \hat{s}}
(v_q^2 + a_q^2) \lambda^{1/2} (M_V^2, M_H^2; \hat{s}) \frac{
\lambda(M_V^2, M_H^2; \hat{s})+12 M_V^2/\hat{s}}{(1-M_V^2/\hat{s})^2} \ ,
\eeq
where the reduced quark couplings to gauge bosons are given in terms of the
electric charge and the weak isospin of the fermion as: $a_q=2I_q^3, v_q=2I_q^3
-4 Q_q s_W^2$ for $V=Z$ and $v_q=-a_q=\sqrt{2}$ for $V=W$, with 
$s_W^2=1-c_W^2\equiv \sin^2\theta_W$.  

The total hadronic cross section is then obtained  by convoluting eq.~(3) with
the parton densities and summing over the contributing partons
\beq
\sigma_{\rm LO} ( pp \to VH)  = \int_{\tau_0}^1  {\rm d}\tau \,
\sum_{q,\bar{q}} \, \frac{ {\rm d} {\cal L}^{q \bar{q}} }{ {\rm d} \tau}
\, {\hat \sigma}_{\rm LO} (\hat{s}= \tau s) \ ,
\eeq
where $\tau_0= (M_V+M_H)^2/s \equiv M_{HV}^2/s$, $s$ being the total hadronic
c.m. energy, and the luminosity is defined in terms of the parton densities
defined at a factorization scale $\mu_F$.\s

In fact, the factorization of the $pp \to HV$ cross section in eq.~(1) holds 
in principle at any order of perturbation theory in the strong interaction and
we can thus  write\footnote{This is only valid at first order in the
electroweak coupling; at two-loop order in $G_F$, QCD corrections to the final
state should also be taken into account. In addition, for the process $pp \to
HZ$, an additional contribution will appear at ${\cal O}(\alpha_s^2)$ as will
be discussed later.}
\beq
\frac{ {\rm d} \hat{\sigma} }{ {\rm d}k^2 } (pp \to H V+X) =  \sigma
(pp \to V^*+X ) \times \frac{ {\rm d} \Gamma }{ {\rm d}k^2 } (V^* \to H V) \ ,
\eeq
where d$\Gamma/$d$k^2$ is given by eq.~(2). Therefore, the  QCD corrections to 
the Higgs-strahlung process, derived at NLO in Refs.~\cite{HVNLO,HVNLOrest}, 
are  simply the corrections to the  Drell--Yan process \cite{DYNLO,DYNNLO}, as 
pointed out in Ref.~\cite{DYequiv}.  \s

At NLO, the QCD corrections to the Drell--Yan process consist of virtual
corrections with gluon exchange in the $q \bar{q}$ vertex and quark self-energy
corrections, which have to be multiplied by the tree-level term, and the
emission of an additional gluon, the sum of which has to be squared and added
to the corrected tree-level term; see Fig.~2. \vspace*{7mm}

\begin{center}
\setlength{\unitlength}{.8pt}
\SetWidth{1.1}
\begin{picture}(180,100)(-50,0)
\ArrowLine(0,100)(50,50)
\ArrowLine(50,50)(0,0)
\Gluon(25,75)(25,25){-3}{5}
\Photon(50,50)(100,50){-3}{5}
\put(85,46){$V^*$}
\put(-10,109){$q$}
\put(-10,4){$\bar{q}$}
\put(10,56){$g$}
\end{picture}
\begin{picture}(180,100)(-30,0)
\ArrowLine(0,100)(50,50)
\ArrowLine(50,50)(0,0)
\Photon(50,50)(100,50){-3}{5}
\GlueArc(27.5,72.5)(12.5,-45,135){3}{4}
\put(85,46){$V^*$}
\put(-10,109){$q$}
\put(-10,4){$\bar{q}$}
\put(40,119){$g$}
\end{picture}
\begin{picture}(180,100)(-10,0)
\ArrowLine(0,100)(50,50)
\ArrowLine(50,50)(0,0)
\Photon(50,50)(100,50){-3}{5}
\Gluon(27.5,72.5)(50,100){-3}{5}
\put(85,46){$V^*$}
\put(-10,109){$q$}
\put(-10,4){$\bar{q}$}
\put(50,90){$g$}
\end{picture}
\end{center}
\centerline{Figure 2: \it NLO QCD corrections to the vector 
boson--quark--antiquark vertex.}
\vspace*{3mm}

Including these contributions, and taking into account the virtuality of the
vector boson, the LO cross section is modified in the following way 
\begin{eqnarray}
\sigma_{\rm NLO} & = & \sigma_{\rm LO} + \Delta\sigma_{q\bar q} +
\Delta\sigma_{qg} \ , 
\eeq
with
\beq
\Delta\sigma_{q\bar q} & = & \frac{\alpha_s(\mu_R)}{\pi} \int_{\tau_0}^1 d\tau
\sum_q \frac{d{\cal L}^{q\bar q}}{d\tau} \int_{\tau_0/\tau}^1 dz~\hat
\sigma_{\rm LO}(\tau z s)~\omega_{q\bar q}(z) \ , \nonumber \\
\Delta\sigma_{qg} & = & \frac{\alpha_s(\mu_R)}{\pi} \int_{\tau_0}^1 d\tau
\sum_{q,\bar q} \frac{d{\cal L}^{qg}}{d\tau} \int_{\tau_0/\tau}^1 dz~\hat
\sigma_{\rm LO}(\tau z s)~\omega_{qg}(z) \ ,
\end{eqnarray}
with the coefficient functions \cite{DYNLO}
\begin{eqnarray}
\omega_{q\bar q}(z) & = & -P_{qq}(z) \log \frac{\mu_F^2}{\tau s}
+ \frac{4}{3}\left[ \left(\frac{\pi^2}{3} -4\right)\delta(1-z) +
2(1+z^2) \left(\frac{\log(1-z)}{1-z}\right)_+ 
\right] \ , \nonumber \\
\omega_{qg}(z) & = & -\frac{1}{2} P_{qg}(z) \log \left(
\frac{\mu_F^2}{(1-z)^2 \tau s} \right) + \frac{1}{8}\bigg[ 1+6z-7z^2 \bigg] \ ,
\end{eqnarray}
where $\mu_R$ denotes  the renormalization scale and $P_{qq}, P_{qg}$  are the
well-known Altarelli--Parisi splitting functions,  which are given by
\cite{apsplit}
\begin{eqnarray}
P_{qq}(z)  =  \frac{4}{3} \left[ \frac{1+z^2}{(1-z)_+}+\frac{3}{2}\delta(1-z)
\right]  \ , \ 
P_{qg}(z) & = & \frac{1}{2} \bigg[ z^2 + (1-z)^2 \bigg] \ .
\end{eqnarray}
The index $+$ denotes the usual distribution $F_+(z)=F(z)-\delta(1-z)\int_0^1
dz' F(z')$. Note that the cross section depends explicitly on $\log(\mu_F^2/Q^2
)$; the factorization scale choice $\mu_F^2 =Q^2 =M_{HV}^2$ therefore avoids
the occurrence of these potentially large logarithms. The  renormalization
scale dependence enters in the argument of $\alpha_s$ and is rather weak. In
most of our discussion, we will set the two scales at $\mu_F= \mu_R= M_{HV}$.
For this choice, the NLO corrections increase the LO cross section by
approximately 30\%.

\subsection*{3. The NNLO corrections}

The NNLO corrections, i.e. the contributions at ${\cal O}(\alpha_s^2)$, to the
Drell--Yan process $pp \to V^*$  consist of the  following set of radiative
corrections [see also Fig.~3a--c]: 

\begin{itemize}
\vspace*{-3mm} 

\item[$a)$] Two-loop corrections to $q\bar{q} \to V^*$, which have to be
multiplied by the Born term.
\vspace*{-3mm}

\item[$b)$] One-loop corrections to the processes $qg \to qV^*$ and $q\bar{q}
\to gV^*$, which have to be multiplied by the tree-level $g q$ and $q\bar{q}$ 
terms initiated by the diagrams shown in Fig.~2.\vspace*{-3mm}

\item[$c)$] Tree-level contributions from $q\bar{q}, qq,qg, gg \to V^*+$ 2 
partons in all possible ways; the sums of these diagrams for a given initial 
and final  state have to be squared and  added. 
\end{itemize}

\begin{center}
\setlength{\unitlength}{1pt}
\SetWidth{1.1}
\begin{picture}(450,100)(-30,0)
\ArrowLine(0,100)(50,50)
\ArrowLine(50,50)(0,0)
\Photon(50,50)(100,50){-3}{5}
\GlueArc(27.5,72.5)(12.5,-45,135){3}{4}
\Gluon(21,20)(21,80){4}{4}
\put(80,35){$V^*$}
\put(-10,10){$q$}
\put(-10,90){$\bar{q}$}
\put(50,-5){${\bf a)}$}
\hspace*{1mm}
\Gluon(170,80)(170,20){4}{5}
\Line(140,80)(210,80)
\Line(140,20)(210,20)
\ArrowLine(210,20)(210,80)
\Gluon(210,20)(260,20){4}{4}
\Photon(210,80)(260,80){-3}{5}
\put(130,80){$q$}
\put(130,20){$\bar{q}$}
\put(170,-5){${\bf b)}$}
\hspace*{1mm}
\ArrowLine(290,80)(340,50)
\ArrowLine(290,20)(340,50)
\Photon(340,50)(380,50){-3}{5}
\Gluon(320,60)(360,80){4}{4}
\Gluon(320,40)(350,20){4}{4}
\put(290,70){$q$}
\put(290,30){$\bar{q}$}
\put(320,-5){${\bf c)}$}
\end{picture} 
\end{center}
\vspace*{2mm}
\centerline{Figure 3: \it Diagrams for the NNLO QCD corrections to the process 
$q\bar{q} \to W^*$.}
\vspace*{3mm}

These corrections have been calculated a decade ago in Ref.~\cite{DYNNLO}
and recently updated \cite{DYupdate}. \s

However, these calculations are not sufficient to obtain a full NNLO
prediction: the cross sections must be folded with the NNLO-evolved parton
distribution funtions (PDFs), which are necessary. The latter require the
calculation of Altarelli--Parisi splitting functions up to three loops and, to
this day, the latter are not completely known at this order. Nevertheless, a
large number of moments of these functions are available \cite{van-neerven};
when these are combined  with additional information on the behaviour at small
$x$, we can  obtain an approximation of the splitting functions at the
required order.  The NNLO MRST \cite{MRST} parton distributions follow this
approach and can therefore be adopted for this calculation. \s

In the case of $ pp \to HZ$ production, because the final state is electrically
neutral, two additional sets of corrections need to be  considered at ${\cal
O}(\alpha_s^2)$.  

Contrary to charged $W$ bosons, the neutral $Z$ bosons can be produced via an
effective $Z$--gluon--gluon coupling induced by quark loops (Fig.~4). This can
occur at  the two-loop level in a box+triangle diagram in  $q\bar{q} \to Z^*$
[which has to be multiplied by the Born term], or at the one-loop level where
vertex diagrams appear for the $q\bar{q} \to gZ^*$ and $qg \to qZ^*$ processes
[and which have to be multiplied by the respective ${\cal O}(\alpha_s)$
tree-level terms]. \s

\begin{center}
\setlength{\unitlength}{1pt}
\SetWidth{1.1}
\begin{picture}(450,100)(-10,0)
\ArrowLine(0,20)(40,20)
\ArrowLine(0,80)(40,80)
\Gluon(40,20)(80,20){4}{4}
\Gluon(40,80)(80,80){4}{4}
\ArrowLine(40,80)(40,20)
\ArrowLine(80,80)(80,20)
\ArrowLine(80,80)(110,50)
\ArrowLine(80,20)(110,50)
\Photon(110,50)(140,50){-3}{5}
\put(120,35){$Z^*$}
\put(58,46){$q$}
\put(-10,18){$q$}
\put(-10,78){$\bar{q}$}
\hspace*{1mm}
\Gluon(180,80)(210,70){4}{4}
\ArrowLine(160,80)(180,80)
\Gluon(160,20)(210,20){4}{4}
\ArrowLine(180,80)(210,99)
\ArrowLine(210,20)(210,70)
\ArrowLine(210,70)(240,50)
\ArrowLine(240,50)(210,20)
\Photon(240,50)(270,50){-3}{5}
\put(150,80){$q$}
\put(150,20){$g$}
\ArrowLine(300,70)(330,50)
\ArrowLine(300,30)(330,50)
\Gluon(330,50)(360,50){4}{4}
\ArrowLine(390,20)(390,80)
\ArrowLine(360,50)(390,80)
\ArrowLine(360,50)(390,20)
\Photon(390,20)(420,20){-3}{5}
\Gluon(390,80)(420,80){4}{3}
\put(290,70){$q$}
\put(290,30){$\bar{q}$}
\end{picture} 
\vspace*{-5mm}
\nn {Figure 4: \it Diagrams for the QCD corrections to  $q\bar{q} \to Z^*$ not
present in $q\bar{q} \to W^*$.} 
\end{center}
\vspace*{3mm}

Because gluons have only vector couplings to quarks and the effective $Zgg$
coupling must be a colour singlet, only the axial--vector part of the
$Zq\bar{q}$ coupling will contribute as a consequence of Furry's theorem. The
axial--vector of the $Z$ coupling to quarks, $a_q = 2I_q^3$, differs only by a
sign for isospin up-type and down-type quarks, so that their contribution
should vanish in the case of quarks that are degenerate in mass. Thus, in the
SM, only the top and bottom quarks will contribute to these
topologies\footnote{Note that  additional contributions involving heavy top
quarks in gluonic two-point  functions are also present; however, they give
zero contribution when the top quark is decoupled.}. \s

These corrections have been evaluated in Refs.~\cite{Scott,Zgg} and have been
shown to be extremely small. The one-loop corrections give relative
contributions that are less than a few times $10^{-4}$ and thus completely
negligible. The  two-loop contribution is somewhat larger. However,  at 
Tevatron energies and for partonic c.m. energies close to the $Z$ boson mass,
the contribution is still below the 1\% level, and it is even smaller at the
LHC \cite{Scott}.  These corrections can therefore be safely neglected. \s

Another set of diagrams that contribute at  ${\cal O}(\alpha_s^2)$ to $ZH$ and
not to $WH$ production [again because of charge conservation] is the
gluon--gluon-initiated mechanism $gg \to HZ$ \cite{ggZH,BK}. It is mediated by 
quark loops [see Fig.~5] which enter in two ways. There is first a triangular
diagram with $gg \to Z^* \to HZ$, in which only the top and bottom quark
contributions are present since,  the  $Z$ boson couples only axially  to the
internal quarks, because of C-invariance, the contribution of a mass-degenerate
quark weak-isodoublet vanishes. There are also box diagrams where both the $H$
and $Z$ bosons are emitted from the internal quark lines and where only the
contribution involving heavy quarks which couple strongly to the Higgs boson  
[the top quark and, to a lesser extent, the bottom quark] are important.

\begin{center}
\setlength{\unitlength}{1pt}
\SetWidth{1.1}
\vspace*{-2mm}
\begin{picture}(450,100)(-10,0)
\Gluon(20,20)(60,20){4}{4}
\Gluon(20,80)(60,80){4}{4}
\ArrowLine(60,20)(60,80)
\ArrowLine(60,80)(100,50)
\ArrowLine(100,50)(60,20)
\Photon(100,50)(150,50){4}{4}
\Photon(150,50)(190,80){4}{4}
\DashLine(150,50)(190,20){5}
\put(120,60){$Z^*$}
\put(68,46){$Q$}
\put(10,18){$g$}
\put(10,78){$g$}
\put(200,20){$H$}
\put(200,80){$Z$}
\hspace*{9cm}
\Gluon(0,20)(40,20){4}{4}
\Gluon(0,80)(40,80){4}{4}
\ArrowLine(40,20)(90,20)
\ArrowLine(40,80)(90,80)
\ArrowLine(40,80)(40,20)
\ArrowLine(90,80)(90,20)
\Photon(90,80)(140,80){4}{4}
\DashLine(90,20)(140,20){5}
\put(145,20){$H$}
\put(58,46){$Q$}
\put(-10,18){$g$}
\put(-10,78){$g$}
\put(145,80){$Z$}
\end{picture} 
\vspace*{-2mm}
\nn {Figure 5: \it Diagrams for the $gg\to HZ$ process, which contributes to 
${\cal O}(\alpha_s^2)$.}
\end{center}

At the LHC, the contribution of this  gluon--gluon fusion mechanism to the $pp
\to HZ$ total production cross section can be substantial. This is due to the
fact that the suppression of the cross section  by a power $(\alpha_s/\pi)^2$
is partly compensated by the increased gluon luminosity  at high-energies. In
addition, the tree-level cross section for $q\bar{q} \to HZ$  drops for
increasing c.m. energy and/or $M_H$ values, since it is mediated by
$s$-channel gauge-boson exchange. \s

We have recalculated the cross section for the process  $gg \to HZ$ at the LHC
energy $\sqrt{s}=14$ TeV, taking into account the full top and bottom quark
mass [$m_b=5$ GeV and $m_t=175$ GeV] dependence. The two contributing triangle
and box amplitudes interfere  destructively, as found in Ref.~\cite{BK}. Our 
results agree with those given in the figures of this article, once we take the
same kinematical configuration, inputs and PDFs [$\sqrt{s}=17$ TeV and
$m_t=80,140$ or 200 GeV].  The cross section for this process is of course
negligible at the Tevatron, because of the low gluon luminosity and the reduced
phase space. 

\subsection*{4. Numerical analysis}

The impact of higher order (HO) QCD corrections is usually quantified by
calculating the $K$-factor, which is defined as the ratio of the cross sections
for the process at HO (NLO or NNLO), with the value of $\alpha_s$ and the PDFs
evaluated also at HO, over the cross section at LO,  with $\alpha_s$ 
and the PDFs consistently  evaluated also at LO:
\beq  
K_{\rm HO}=\frac{\sigma_{\rm HO}  (pp \to HV+X) }{\sigma_{\rm LO}( pp \to HV) } 
\ . 
\eeq 
A kind of $K$-factor for the LO cross section, $K_{\rm LO}$, can also be 
defined by  evaluating the latter at given factorization and renormalization
scales and normalizing to the LO cross sections evaluated at the central scale, 
which, in our case, is given by $\mu_F=\mu_R=M_{HV}$.\s

The $K$-factors at NLO and NNLO are shown in Figs.~6 and 7 (solid black lines) 
for, repectively, the LHC and   the Tevatron  as a function of the Higgs  mass
for the process $pp \to HW$; they are practically the same\footnote{ Because of
the slightly different phase space and scale, the $K$-factor for $pp \to ZH$ is
not identical to the $K$-factor for $pp \to WH$. However, since
$(M_Z^2-M_W^2)/\hat{s}$ is small and the dependence  of d$\Gamma$ in eq.~(3) on
$k^2$ is not very strong in the range that we are considering, the $K$-factors
for the two processes are very  similar.} for the process $pp \to HZ$ when the
contribution of the $gg \to HZ$ component is not included. The  scales have
been fixed to $\mu_F=\mu_R=M_{HV}$ and the MRST sets of PDFs for each
perturbative order are used in a consistent manner.\s

The NLO $K$-factor is practically constant at the LHC, increasing only from
$K_{\rm NLO}=1.27$ for $M_H=110$ GeV to $K_{\rm NLO}=1.29$ for $M_H=300$ GeV.
The NNLO contributions increase the $K$-factor by a mere 1\% for the low $M_H$
value and by 3.5\% for the high value. At the Tevatron, the NLO $K$--factor is 
somewhat higher than at the LHC, enhancing the cross section between  $K_{\rm
NLO}=1.35$ for $M_H=110$ GeV and $K_{\rm NLO}=1.3$ for $M_H=300$ GeV with a
monotonic decrease.  The NNLO corrections increase the $K$-factor uniformly by
about 10\%. Thus, these NNLO corrections are more important at the Tevatron
than at the LHC. \s

The bands around the $K$-factors represent the variation of the cross sections
when they are evaluated at renormalization and factorization scale values that
are independently varied from $\frac{1}{3} M_{HV} \leq \mu_F \, (\mu_R) \leq 3
M_{HV}$, while the other is fixed to $\mu_R \, (\mu_F) = M_{HV}$; the
normalization provided by the production cross section evaluated at scales
$\mu_F=\mu_R=M_{HV}$. As can be seen, except from the accidental cancellation
of the scale dependence of the LO cross section at the LHC, the decrease of the
scale variation is strong when going from LO to NLO and then to NNLO. For
$M_H=120$ GeV, the uncertainty from the scale choice at the LHC drops from 10\%
at LO, to 5\% at NLO, and to 2\% at NNLO. At the Tevatron and for the same
Higgs boson mass, the scale uncertainty drops from 20\% at LO, to 7\% at NLO,
and to 3\% at NNLO. If this variation of the cross section with the two scales
is taken as an indication of the uncertainties due to the not yet calculated
higher order corrections, one concludes that once the NNLO contributions are
included in the prediction, the cross section for the $pp \to HV$ process is
known at the rather accurate level of 2 to 3\%.  \s

Finally,  we present in Fig.~7 the total production cross sections at NNLO for
the processes $q\bar{q} \to HW$ and $HZ$ at both the Tevatron and the LHC as a
function of $M_H$. In the case of the $HZ$ process, the contribution of the $gg
\to ZH$ subprocess to the total cross section is not yet included, but it is
displayed separately in the  LHC case. For Higgs masses in the range 100  GeV 
$\lsim M_H \lsim 250$ GeV, where $\sigma(q\bar{q} \to HZ)$ is significant,
$\sigma(gg \to HZ)$ is at the level of 0.1 to 0.01 pb and represents about 10\%
of the total cross section for low $M_H$. 

\begin{figure}[htbp]
\begin{center}
\psfig{figure=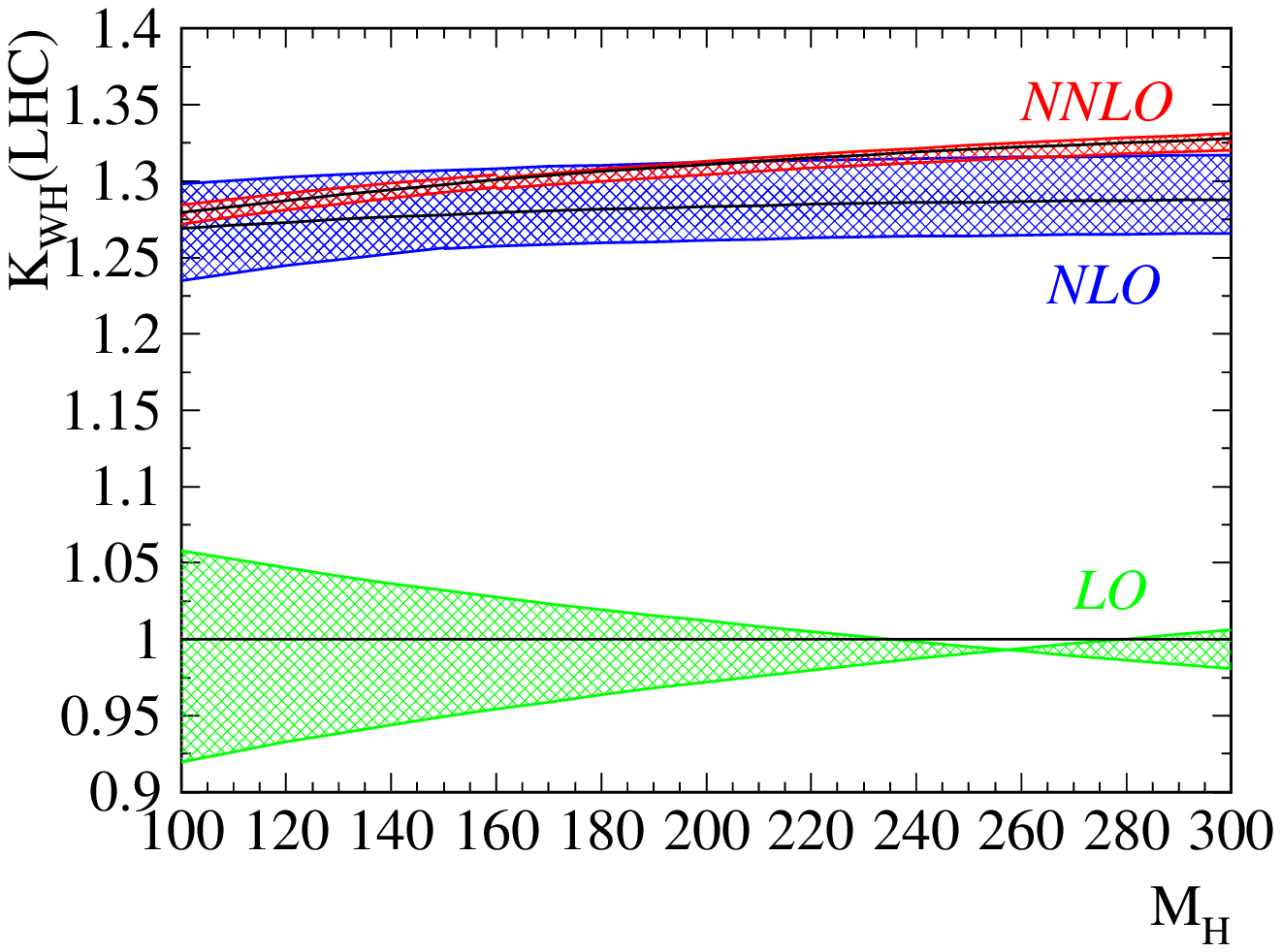,bbllx=50,bblly=300,bburx=560,bbury=530,width=17cm}
\vspace*{0.5cm}
\end{center}
{Figure 6:  \it The $K$-factors for $pp \to HW$ at the LHC  as  a function of
$M_H$ at LO, NLO and NNLO (solid black lines). The bands represent the spread
of the cross section when the renormalization and factorization scales are
varied in the range $\frac{1}{3}M_{HV} \leq \mu_R\, (\mu_F) \leq 3M_{HV}$, the
other scale being fixed at $\mu_F (\mu_R)= M_{HV}$.}
\vspace*{0mm}
\end{figure}

\begin{figure}[htbp]
\begin{center}
\psfig{figure=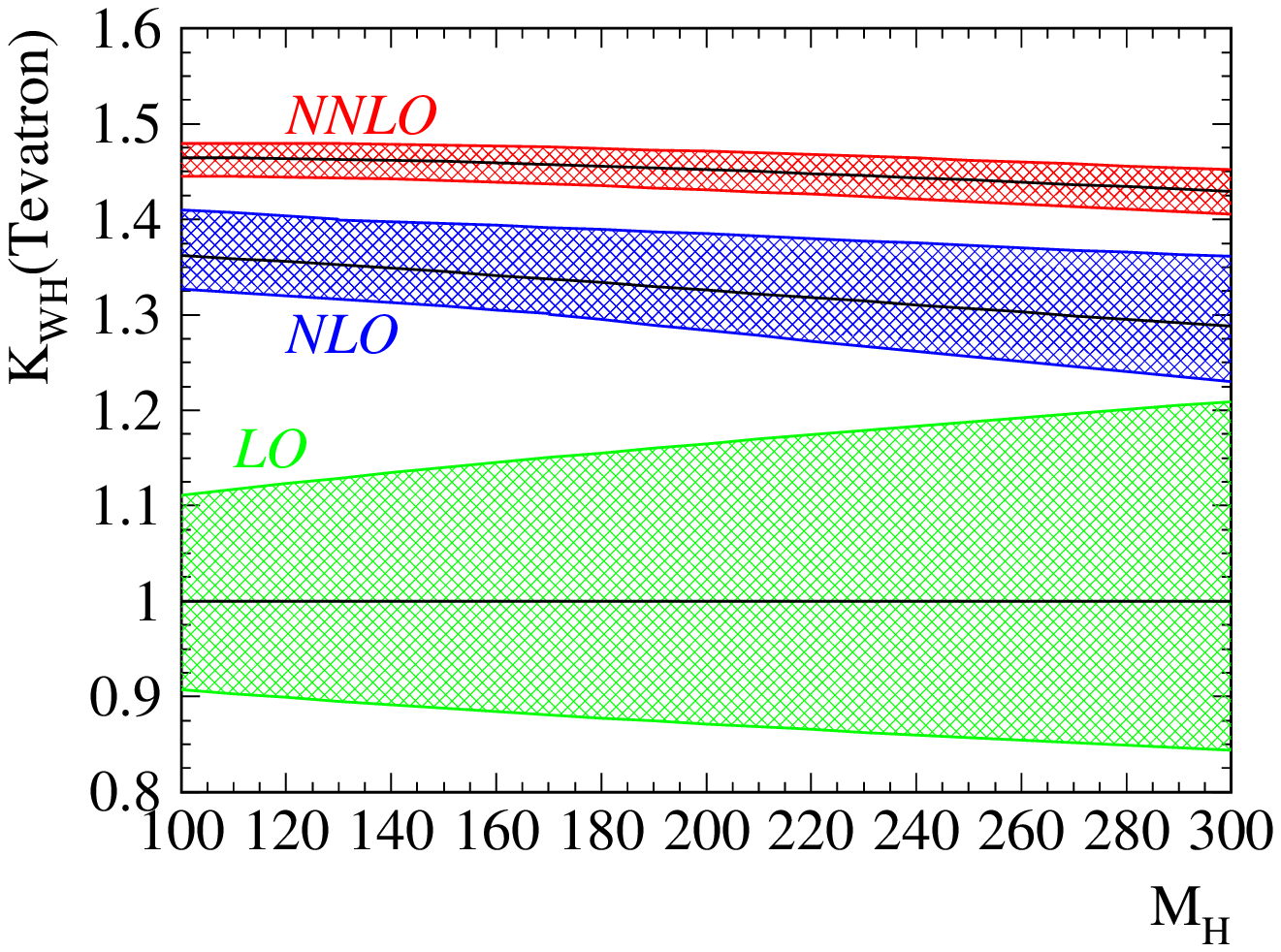,bbllx=50,bblly=300,bburx=560,bbury=530,width=17cm}
\end{center}
\vspace*{.7cm}
\centerline{Figure 7:  \it The same as in Fig.~6 but for the Tevatron case,
$p\bar{p} \to HW$.}
\end{figure}

Its relative magnitude increases  for higher Higgs masses; for very large $M_H$
values, it reaches the level of the Drell--Yan cross section.  However, for
these large Higgs masses, the total production rate is too small to be useful
in practice.   Note that at the Tevatron,  as expected, the cross section of
the  $gg \to HZ$ subprocess is very small, barely reaching the level  of
$\sigma(gg \to  HZ) \sim 0.2$ fb for $M_H=120$ GeV. The contribution of this
subprocess can be safely neglected in this case.  

\begin{figure}[h!]
\begin{center}
\vspace*{-2.5cm}
\hspace*{-2cm}
\psfig{figure=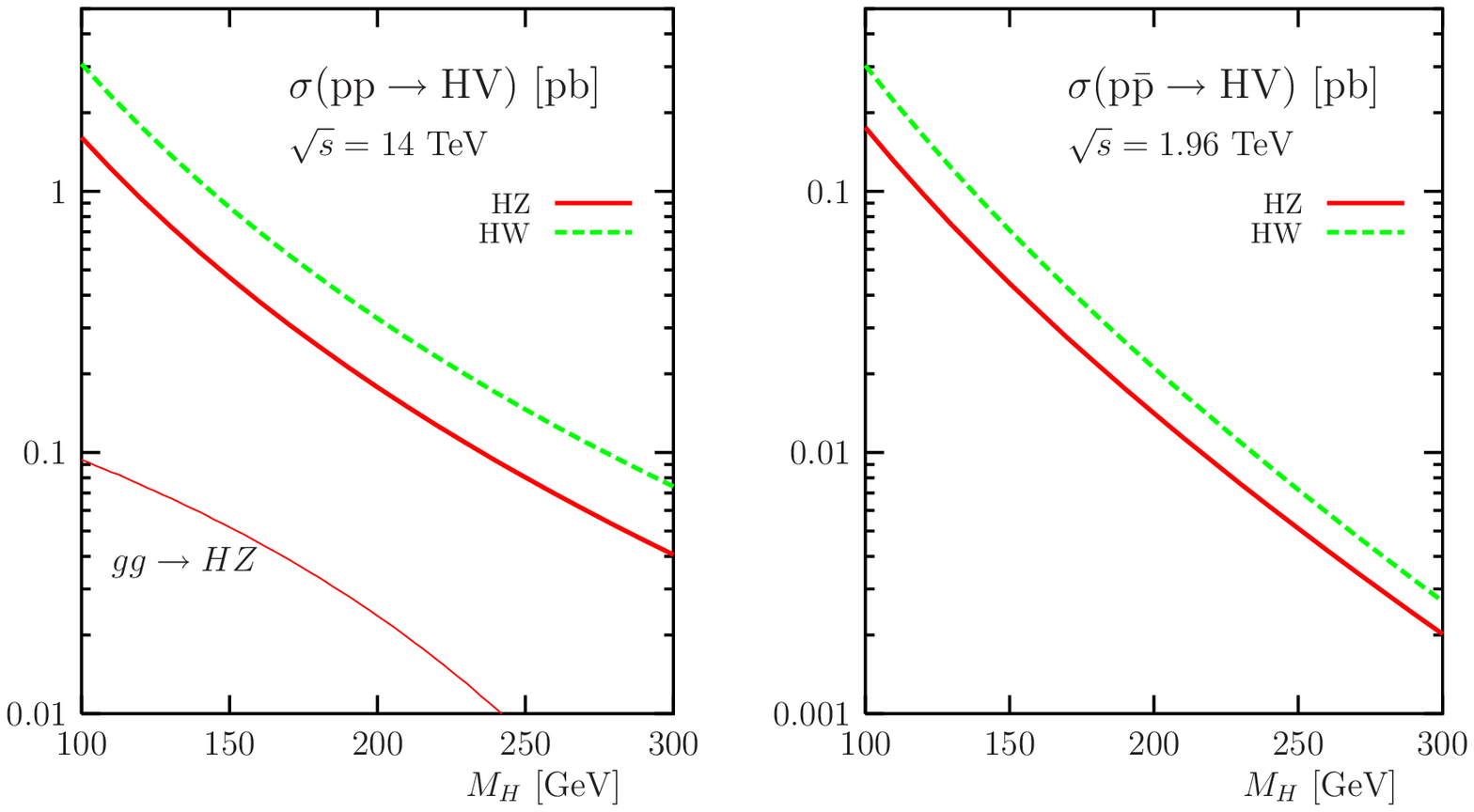,width=19.5cm}
\vspace*{-17.cm}
\end{center}
{\it Figure 8:   The total production cross sections at NNLO for $pp \to
HW$ and $HZ$ at the  LHC (left) and the Tevatron (right) as a
function of $M_H$. The MRST  parton densities have been
used. The contribution of the $gg \to HZ$ process is shown separately in the
case of the LHC.}
\vspace*{-5mm}
\end{figure}

\subsection*{5. Concluding remarks}

We have discussed the Higgs-strahlung processes $pp \to HV$ with $V=W,Z$ at
NNLO in strong interactions. We have shown that the Drell--Yan-type corrections
increase the total production cross sections by up to 3\% at the LHC and by up 
to 10\% at the Tevatron in the Higgs mass range relevant at these colliders.
Because of the larger gluon luminosity at high energies, the additional
contribution due to the $gg \to HZ$ subprocess can be relatively important at
the LHC, and for Higgs boson masses far above 200 GeV, it becomes comparable  
with the $q\bar{q} \to HZ$ production cross section, but the total rate is 
then rather small. The scale dependence is strongly reduced from NLO,  where it can
reach a level close to 10\%, to less than 2--3\% at NNLO. These NNLO
corrections are of the same order, but of opposite sign, as the ${\cal
O}(\alpha)$ electroweak corrections to these processes,  which have been
calculated very recently \cite{EWcor}.\s

Together with the effects of higher-order corrections, the uncertainties due to
the PDFs dominate the theoretical error on  the production cross section.
Recently, the CTEQ \cite{CTEQ6} and MRST \cite{MRST2001E} collaborations
introduced a new  scheme, which gives the possibility of controlling these
errors: in addition to the nominal best fit PDFs, they provided a set of $2N$
PDFs $[N=20$ for CTEQ and $15$ for MRST] at NLO  [they are not yet available at
NNLO], corresponding to the minima and maxima of the $N$ eigenvectors of the 
matrix error of the fitting parameters. Adding the maximum and minimum
deviation for each eigenvector in quadrature, we obtain an error on the total
cross section at NLO of less than 5\% for $M_H \lsim 300$ (200) GeV for
LHC (Tevatron) energies \cite{Samir}. \s

All these features make Higgs-strahlung one of the most theoretically clean
Higgs boson production channels at hadron colliders. This will be of great
importance when it comes to the determination of the properties of the Higgs 
boson and the measurement of its couplings. The systematic uncertainties
originating from higher-order corrections and structure functions  being small
[a better determination of the parton distribution functions can be performed
in the future], the Higgs-strahlung process will provide a clean determination
of the $HVV$ couplings times the Higgs branching ratios [the latter being
measured in other Higgs production processes] if enough integrated luminosity
is collected to make the statistical errors  also small\footnote{Note that an
additional systematic error of about 5\% arises from the $pp$ luminosity.  To
reduce all these uncertainties at the LHC, it has been suggested \cite{Michael}
to use the Drell--Yan processes $pp \to W,Z$ with the subsequent leptonic
decays of the gauge bosons as a means of measuring directly the quark and
antiquark luminosities at hadron colliders. The errors on the cross sections
for all processes dominated by $q\bar{q}$ scattering, such as the 
Higgs-strahlung discussed here, when normalized to the Drell--Yan rate, would
lead to a total  systematical uncertainty of less than 1\%. In this case, the
dominant part of the $K$-factor for Higgs-strahlung  will drop out in the
ratio.}. \s

This analysis can straightforwardly be extended to the case of the  Minimal
Supersymmetric extension of the SM \cite{HHG}. The two CP-even Higgs bosons,
$\Phi=h$ and $H$, can be produced in the same channels, $pp \to V\Phi$, and 
the LO cross sections are the same as the one for the SM Higgs boson, except
that they are suppressed by global coupling factors, $0 \leq g_{\Phi VV}^2
\leq  1$. The standard QCD corrections are again similar to those discussed
here; the additional corrections due to supersymmetric partners of quarks and 
gluons have to be added [note that the triangular $Zgg$ diagrams with scalar
loops give zero contribution here]. This analysis can even be  extended to the
case of the associated production of CP-even and CP-odd Higgs particles, $pp
\to \Phi A$. Here again, the cross section can be factorized into the product
of $pp \to V^*$ production and $V^* \to \Phi A$ decays, and  the QCD
corrections are thus the same as those discussed here. A small additional
contribution originating from the one-loop subprocess $gg \to \Phi A$
\cite{trilin} has to be added, though.  \bigskip

\nn {\bf Acknowledgements:} We thank Samir Ferrag and Michael Spira for
discussions.

\end{document}